# Statistics of the dissipated energy in driven diffusive systems

A. Lasanta[1,a], Pablo I. Hurtado[2,3,b], and A. Prados[4,c]

[1] CNR-ISC and Dipartimento di Fisica, Università La Sapienza, p.le A. Moro 2, 00185 Rome, Italy
[2] Instituto Carlos I de Física Teórica y Computacional, Universidad de Granada, Granada 18071, Spain
[3] Departamento de Electromagnetismo y Física de la Materia, Universidad de Granada, Granada 18071, Spain
[4] Física Teórica, Universidad de Sevilla, Apartado de Correos 1065, E-41080 Sevilla, Spain

**Abstract.** Understanding the physics of non-equilibrium systems remains as one of the major open questions in statistical physics. This problem can be partially handled by investigating macroscopic fluctuations of key magnitudes that characterise the non-equilibrium behaviour of the system of interest; their statistics, associated structures and microscopic origin. During the last years, some new general and powerful methods have appeared to delve into fluctuating behaviour that have drastically changed the way to address this problem in the realm of diffusive systems: macroscopic fluctuation theory (MFT) and a set of advanced computational techniques that make it possible to measure the probability of rare events. Notwithstanding, a satisfactory theory is still lacking in a particular case of intrinsically non-equilibrium systems, namely those in which energy is not conserved but dissipated continuously in the bulk of the system (e.g. granular media). In this work, we put forward the dissipated energy as a relevant quantity in this case and analyse in a pedagogical way its fluctuations, by making use of a suitable generalisation of macroscopic fluctuation theory to driven dissipative media.

[a] e-mail: `alasanta@us.es`
[b] e-mail: `phurtado@onsager.ugr.es`
[c] e-mail: `prados@us.es`

# 1 Introduction

Non-equilibrium systems are ubiquitous in nature. During the last years, the statistical physics community has devoted an enormous effort on the search of a general theory of non-equilibrium media, akin to the existing ensemble theory for equilibrium systems. Recently, the study of fluctuation statistics of macroscopic observables in the realm of stationary non-equilibrium states has apparently opened the door to such a general theoretical framework in the form of a macroscopic fluctuation theory (MTF) [1], built upon the language of large deviations. Interestingly, in the case of equilibrium systems, this approach provides an alternative way to derive thermodynamic potentials [3].

Within this scheme it is of crucial importance to choose the macroscopic observable to study. For example, in driven diffusive systems some quantity is typically conserved, such as density, energy, momentum etc. This suggests that the relevant observable to be considered here is the associated current or flux that traverses the system when driven by boundary gradients or external fields. In this way, the large deviation function (LDF) that describes the statistics of current fluctuations seems to play a role similar to the equilibrium free energy [2]. Indeed, large deviation theory offers a general mathematical framework to describe both equilibrium and non-equilibrium statistical mechanics [4].

Much less studied is a paradigmatic class of intrinsically non-equilibrium systems: dissipative ones. In this broad class of systems energy is continuously dissipated in the bulk, so a continuous input of energy is needed in order to reach and sustain a stationary state. A typical example is that of in granular media: when two particles collide, a certain fraction of their energy is lost to the environment [5]. It is therefore no surprise that the dissipated energy is one of the main quantities to study when developing their statistical mechanics description.

The broad class of driven dissipative systems does not only include granular media, but also dissipative electronics, chemical-reactions systems and, in general, all reaction-diffusion systems in which dissipation, diffusion and driving are the main ingredients. One of their main common physical features is that local gradients may appear even in the absence of boundary driving, with a characteristic length that depends on the dissipation [5]. The main aim of this paper is to present a general and pedagogical overview of a generalisation of MFT to this large class of dissipative systems, recently introduced by the authors [7,8]. In particular, we focus below on how to calculate the LDF of the dissipated energy, as the key relevant observable for this family of non-equilibrium systems.

The plan of the paper is as follows. In section 2 we describe the fluctuating balance equation ruling the mesoscopic dynamics of a general class of diffusive systems with dissipation, and we highlight the definition of a paradigmatic microscopic model within this family. The generalisation of MFT to driven dissipative media is developed in section 3. In the following section 4, we consider the general constrained variational problem that stems from the MFT when the probability of the fluctuation of the space- and time-integrated dissipation is considered. The specific case of a system thermostatted at the boundaries is then analysed throughout the following sections: first, the relevant boundary conditions are derived in 4.1; second, the mapping onto an unconstrained variational problem with a second-order-derivative Lagrangian is discussed in 5 and, finally, an equivalent Hamiltonian description is presented in 5.1 and put at work in a simpler, weak-dissipation limit. Numerical methods to measure the large deviation statistics of the dissipated energy in Monte Carlo simulations of microscopic models, and to solve numerically the complex equations of MFT, are described in some detail in section 6. The paper ends with its main conclusions in section 7, and some technical notes and a specific code to solve the equations of MFT in some cases is included in an appendix for the benefit of the reader.

## 2 Diffusive systems with dissipation: fluctuating energy balance and microscopic models

In this paper we consider a broad class of dissipative systems of technological and theoretical interest whose dynamics is described at the mesoscopic level by the following balance equation

$$\partial_t \rho(x,t) = -\partial_x j(x,t) + \mathrm{d}(x,t). \tag{1}$$

Here, $\rho(x,t)$ is the local density of the magnitude of interest that, for the sake of concreteness, we consider to be the local energy of the system throughout this work. Moreover, $j(x,t)$ is the energy current and $\mathrm{d}(x,t) < 0$ is the dissipation field. In the previous mesoscopic evolution equation, the macroscopic space and time variables, $t$ and $x \in [-1/2, 1/2]$, arise after a diffusive scaling limit such that $x = \tilde{x}/L$ and $t = \tilde{t}/L^2$, with $\tilde{x}$ and $\tilde{t}$ the microscopic space and time variables and $L$ the system length[1]. Note that this scaling may also involve a continuum limit of the underlying microscopic dynamics [6], see below.

The total energy per particle at time $t$ can be now calculated by integrating $\rho(x,t)$ over space

$$e(t) = \int_{-1/2}^{1/2} dx \, \rho(x,t). \tag{2}$$

The presence of dissipation in Eq. (1) makes total energy a non-conserved quantity,

$$\frac{d}{dt} e(t) = \int_{-1/2}^{1/2} dx \, \mathrm{d}(x,t). \tag{3}$$

It is important to note that the density, current and dissipation fields ($\rho$, $j$ and $\mathrm{d}$, respectively) are fluctuating quantities. In particular, for a general class of systems it can be shown that the energy current obeys

$$j(x,t) = -D(\rho)\partial_x \rho(x,t) + \xi(x,t), \tag{4}$$

where $D(\rho)$ is the diffusivity and $\xi(x,t)$ is the current noise. That is, the energy current is given by Fourier's law plus a fluctuating term. In general, $D(\rho)$ above is a nonlinear function of the local density, whereas the noise $\xi$ is expected to be Gaussian and white: The microscopic interactions in a certain system might be tremendously complicated, but the fluctuations of the slow hydrodynamic fields stem from the sum of a very large number of random events. Therefore, as a consequence of the central limit theorem, Gaussian statistics emerge with an amplitude of the order of $L^{-1/2}$ in the same mesoscopic limit in which (1) is valid. In this way

$$\langle \xi(x,t) \rangle = 0, \qquad \langle \xi(x,t)\xi(x',t') \rangle = \frac{\sigma(\rho)}{L}\delta(x-x')\delta(t-t'), \tag{5}$$

where $\sigma(\rho)$ -which is also a nonlinear function of the local density- is the mobility transport coefficient. The dissipation field is also subject to fluctuations,

$$\mathrm{d}(x,t) = -\nu R(\rho) + \eta(x,t), \tag{6}$$

---

[1] We take $x \in [-1/2, 1/2]$ in order to make it easier to introduce symmetry considerations whenever possible, see below.

where $\nu$ is the macroscopic dissipation coefficient, $R(\rho)$ is another transport coefficient which depends non-linearly on the local density, and $\eta(x,t)$ is the dissipation noise. Similarly to the current noise, we expect the dissipation noise to be Gaussian and white at the mesoscale, on the same physical grounds (this is indeed the case in a general class of diffusive systems with dissipation considered in [6–8])

$$\langle \eta(x,t) \rangle = 0, \qquad \langle \eta(x,t)\eta(x',t') \rangle = \nu^2 \frac{\kappa(\rho)}{L^\beta} \delta(x-x')\delta(t-t'), \tag{7}$$

where $\kappa(\rho)$ is the amplitude of the dissipation noise and $\beta$ is an exponent that depends on the relation between the macroscopic and microscopic dissipation coefficients, see below and also [6]. If $\beta = 1$, both the current and dissipation noises scale as $L^{-1}$ and are equally important to investigate the fluctuations. On the other hand, if $\beta > 1$, the dissipation noise is subdominant against the current noise and can be neglected. As we will argue below, the latter is the relevant situation for systems where diffusion and dissipation compete on the same timescale.

Before continuing with the macroscopic description of fluctuations in these systems, let us briefly introduce a broad class of dissipative lattice models with stochastic microscopic dynamics whose mesoscopic evolution equation corresponds to Eqs. (1) and (4). These models contain the essential ingredients characterising many dissipative media, namely nonlinear diffusive dynamics, bulk dissipation, and boundary driving, and therefore constitute the ideal test-bed where to check our extension of MFT to driven dissipative media. For the sake of simplicity, we describe below a one-dimensional (1D) example with constant diffusivity, but the generalisation to arbitrary dimension and non-linear transport coefficients is straightforward [6–8]. Our model system is thus defined on a 1D lattice of size $L$, where each site $i \in [1, L]$ is characterised by an energy $\rho_i$. The dynamics is stochastic and proceeds via collisions between randomly chosen of nearest neighbours $(i, i+1)$, with total energy $E_i \equiv \rho_i + \rho_{i+1}$, such that a fraction

$$D_i \equiv (1-\alpha)E_i \tag{8}$$

of the pair energy is dissipated out the system, while the remaining energy $\alpha E_i$ is randomly redistributed within the pair. Here $\alpha$ is a constant positive parameter playing the role of a restitution coefficient, $0 < \alpha \le 1$, and note that for $\alpha = 1$ we recover the (conservative) Kipnis-Marchioro-Presutti model of heat conduction [9]. In this way, the energies of the colliding pair after the interaction are

$$\begin{aligned} \rho_i' &= z\alpha E_i \\ \rho_{i+1}' &= (1-z)\alpha E_i, \end{aligned} \tag{9}$$

being $z$ a random number uniformly distributed in $[0, 1]$. In order to drive the system to a stationary state, we couple boundary sites $i = 1, L$ to thermal baths at fixed temperatures $T_r$ (right) and $T_l$ (left). In this way, every time a boundary pair is randomly chosen to collide, the new energy of the boundary site is

$$\rho_{1,L}' = z(\tilde{\rho}_{r,l} + \rho_{1,L}) \tag{10}$$

with $\tilde{\rho}_{r,l}$ an energy randomly drawn from a Boltzmann distribution at the corresponding temperature $T_{r,l}$, i. e. with probability $P(\tilde{\rho}_{r,l}) = T_{r,l}^{-1} \exp(-\tilde{\rho}_{r,l}/T_{r,l})$ (our units are such that $k_B = 1$). This system, which can be considered as an simplified model of a granular media, hence reaches a steady state where boundary energy injection and bulk dissipation balance each other. It is then easy to show that, at the mesoscopic scale, the evolution equation governing the model dynamics corresponds to Eqs. (1)-(7), with a constant diffusivity $D(\rho) = 1/2$, a mobility $\sigma(\rho) = \rho^2$ and a dissipation transport coefficient $R(\rho) = \rho$. Furthermore, this model can be generalised by defining energy-dependent collision rates, in which case a whole class of non-linear driven dissipative systems is obtained. For a detailed description see ref. [6].

## 3 Generalisation of macroscopic fluctuation theory to diffusive systems with dissipation

In this Section we briefly describe how to generalise the macroscopic fluctuation theory (MFT) developed by Bertini et al. [1] to dissipative systems. Our first aim is to write down an explicit expression for the probability of a certain history $\{\rho, j, d\}_0^\tau$ of the relevant fields during a time interval $(0, \tau)$. As described in previous section, in the most general situation both the current and dissipation noises must be taken into account to write the functional associated to this probability. Nevertheless we will argue that, with a great generality, the underlying microscopic dynamics that gives rise in some continuum limit to the mesoscopic equation (1) is quasi-elastic, and that this quasi-elasticity implies that the dissipation noise is subdominant, that is, $\beta > 1$ in (7). In few words, dynamics must be quasi-elastic in order for diffusion and dissipation to compete on equal footing at large spatial and temporal scales. More in detail, we have seen in the model introduced in the previous section that the energy dissipated in a collision is proportional to $1 - \alpha$, with $\alpha$ a proxy of the restitution coefficient of granular media [6, 8, 10]. Of course, the macroscopic dissipation coefficient $\nu$ in Eq. (6) is then proportional to $1 - \alpha$. Quasi-elasticity thus means in this context that $1 - \alpha$ scales as $L^{-2}$, and this can be shown to lead to a reaction-diffusion Langevin equation[2] similar to Eq. (1). This quasi-elasticity of microscopic dynamics then implies that the dissipation noise is subdominant when compared to the current noise, due to a factor $(1 - \alpha)$ in the dissipation field [6, 8, 10]. Therefore, in the remainder of the paper, we will simply write

$$d(x, t) = -\nu R(\rho(x, t)),\tag{11}$$

so that the fluctuations of the dissipation field are enslaved to those of the density field, and only the current noise must be taken into account for the path integral formalism leading to the MFT action below.

Following the discussion of the previous paragraph, the probability of a history $\{\rho, j\}_0^\tau$ in the time interval $(0, \tau)$ is given by

$$P(\{\rho, j\}_0^\tau) \sim \exp\left(-L\, \mathcal{S}[\rho, j]\right),\tag{12}$$

with a rate functional or *action*

$$\mathcal{S}[\rho, j] = \int_0^\tau dt \int_{-1/2}^{1/2} dx\, \frac{[j + D(\rho)\partial_x \rho]^2}{2\sigma(\rho)}\tag{13}$$

with $\rho(x, t)$ and $j(x, t)$ coupled via the balance equation (1), and the dissipation $d(x, t)$ enslaved to the density field $\rho(x, t)$ by (11). The above expression simply follows from the Gaussian character of the current noise [1, 3]. Any observed fluctuation is built up, in some sense, as a certain superposition of these fundamental local Gaussian fluctuations. At first sight, the functional in Eqs. (12)-(13) is equal to that of the conservative ($\alpha = 1$) case, but new physics arises due to the relation (11) between the dissipation and the density fields. In fact, we will see below that, even in the simplest situation, the associated variational problem is considerably more complex than in the conservative case, involving a *Lagrangian* with second order derivatives.

---

[2] In the microscopic dynamics, the diffusion term includes a second spatial derivative, which gives a $L^{-2}$ scaling factor when the mesoscopic space variable $x = \tilde{x}/L$ is defined in terms of the microscopic space variable $\tilde{x}$. On the other hand, the dissipation term is proportional to $(1 - \alpha)$ and has no spatial derivatives, therefore $1 - \alpha$ must scale as $L^{-2}$ in order to have dissipation acting over the same space and time scale as diffusion.

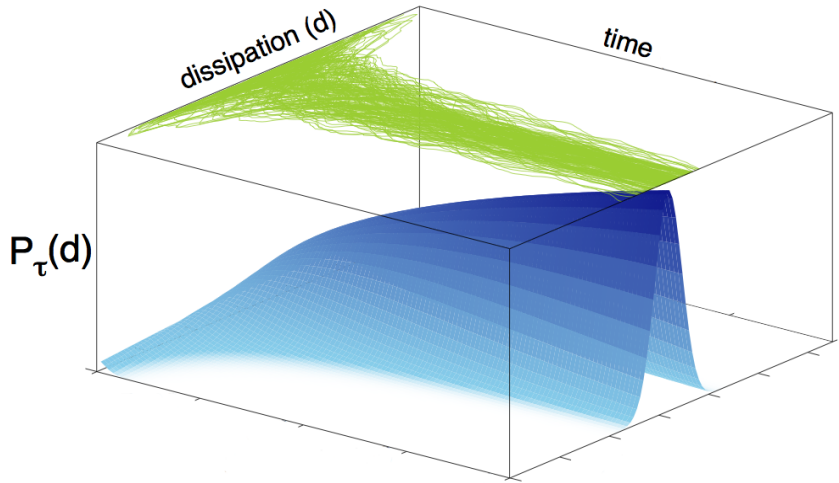

**Fig. 1.** (Color online) Sketch of the convergence of the space- and time-integrated dissipation to its ensemble average value for many different trajectories of the microscopic dynamics. This plot corresponds to a specific case of the class of dissipative systems described in Section 2 [8]. The probability concentrates exponentially fast as time increases, a signature of the validity of a large deviation principle, as given by Eq. (17).

Clearly, a first relevant macroscopic quantity capturing the physics of driven dissipative media is the amount of energy dissipated in the time interval $(0, \tau)$. We define the integrated dissipation $d$ as

$$d = -\frac{1}{\tau} \int_0^\tau dt \int_{-1/2}^{1/2} dx \, \mathrm{d}(x, t) = = \frac{\nu}{\tau} \int_0^\tau dt \int_{-1/2}^{1/2} dx \, R(\rho(x, t)) > 0 \,. \qquad (14)$$

Next, we look into the fluctuations of $d$: in order to calculate the probability $\mathrm{P}_\tau(d)$ of a certain value of the integrated dissipation $d$, we have to sum the probabilities corresponding to all the different histories $\{\rho, j\}$ that lead thereto, that is,

$$\mathrm{P}_\tau(d) = \int \mathcal{D}\rho \, \mathcal{D}j \, \mathrm{P}\left(\{\rho, j\}_0^\tau\right) \, \delta\left(d - \frac{1}{\tau} \int_0^\tau dt \int_{-1/2}^{1/2} dx \, d(x, t)\right) \,, \qquad (15)$$

with the path integral restricted to density and current fields coupled at every point of space and time by the balance equation

$$\partial_t \rho + \partial_x j + \nu R(\rho) = 0 \,, \qquad (16)$$

with adequate boundary conditions for the physical situation of interest. In the large system size limit $L \gg 1$, the above path integral over the different histories $\{\rho, j\}_0^\tau$ is dominated by its saddle point contribution. Therefore, the dissipation distribution obeys a large deviation principle [4]

$$\mathrm{P}_\tau(d) \sim \exp\left[+\tau L \, G(d)\right] \,, \qquad (17)$$

see Fig. 3, with the dissipation large deviation function (LDF) given by the following variational problem

$$G(d) = -\frac{1}{\tau} \min_{\{\rho, j\}} \mathcal{S}[\rho, j],\tag{18}$$

subject to the constraints (14) and (16) and the imposed boundary conditions. From a physical point of view the optimal fields solution of this variational problem, denoted here as $\rho_0(x, t; d)$ and $j_0(x, t; d)$, are the paths the system follows in mesoscopic phase space in order to sustain a given fluctuation of the integrated dissipation $d$.

## 4 The constrained variational problem

The problem described in the previous section is a classical variational problem with constraints, e.g. as those found in classical mechanics. To make clear this parallelism, let us define a *Lagrangian* $\mathcal{L}$,

$$\mathcal{L}(\rho, j, \partial_x \rho) = \frac{[j + D(\rho)\partial_x \rho]^2}{2\sigma(\rho)},\tag{19}$$

such that

$$G(d) = -\frac{1}{\tau} \min_{\rho, j} \mathcal{S}[\rho, j], \qquad \text{with } \mathcal{S}[\rho, j] \equiv \int_0^\tau dt \int_{-1/2}^{1/2} dx \, \mathcal{L}(\rho, j, \partial_x \rho),\tag{20}$$

and subject to the constraints Eq. (14) on the integrated dissipation and Eq. (16) on the balance equation, together with some yet-unspecified boundary conditions. In order to deal with these constraints, we use standard variational calculus techniques [20] and introduce two Lagrange multipliers, a function $f(x, t)$ associated to the balance equation (16) and a parameter $\lambda$ associated to the desired integrated dissipation value in (14). Thus, we seek the minimum of the modified action

$$\tilde{\mathcal{S}}[\rho, j, f; \lambda] = \int_0^\tau dt \int_{-1/2}^{1/2} dx \, \tilde{\mathcal{L}}(\rho, j, f, \partial_t \rho, \partial_x \rho, \partial_x j; \lambda),\tag{21}$$

$$\tilde{\mathcal{L}}(\rho, j, f, \partial_t \rho, \partial_x \rho, \partial_x j; \lambda) \equiv \mathcal{L}(\rho, j, \partial_x \rho) + f(x, t) \left[ \partial_t \rho + \partial_x j + \nu R(\rho) \right] + \lambda \left( \nu R(\rho) - \tau d \right).\tag{22}$$

The fields $\{\rho_0(x, t), j_0(x, t), f_0(x, t)\}$ and the parameter $\lambda_0$ which give the minimum are the solution of the Euler-Lagrange equations

$$\partial_t \left( \frac{\partial \tilde{\mathcal{L}}}{\partial(\partial_t \rho)} \right) + \partial_x \left( \frac{\partial \tilde{\mathcal{L}}}{\partial(\partial_x \rho)} \right) - \frac{\partial \tilde{\mathcal{L}}}{\partial \rho} = 0,\tag{23}$$

$$\partial_t \left( \frac{\partial \tilde{\mathcal{L}}}{\partial(\partial_t j)} \right) + \partial_x \left( \frac{\partial \tilde{\mathcal{L}}}{\partial(\partial_x j)} \right) - \frac{\partial \tilde{\mathcal{L}}}{\partial j} = 0.\tag{24}$$

$$\partial_t \left( \frac{\partial \tilde{\mathcal{L}}}{\partial(\partial_t f)} \right)^{0} + \partial_x \left( \frac{\partial \tilde{\mathcal{L}}}{\partial(\partial_x f)} \right)^{0} - \frac{\partial \tilde{\mathcal{L}}}{\partial f} = 0.\tag{25}$$

and the extremum condition

$$\partial_\lambda \tilde{\mathcal{L}} = 0.\tag{26}$$

Now we use (22) to write

$$\partial_t \cancel{\left(\frac{\partial \mathcal{L}}{\partial(\partial_t \rho)}\right)}^0 + \partial_x \left(\frac{\partial \mathcal{L}}{\partial(\partial_x \rho)}\right) + \partial_t f = \frac{\partial \mathcal{L}}{\partial \rho} - f \nu R'(\rho) + \lambda \nu R'(\rho), \qquad (27)$$

$$\partial_t \cancel{\left(\frac{\partial \mathcal{L}}{\partial(\partial_t j)}\right)}^0 + \partial_x \cancel{\left(\frac{\partial \mathcal{L}}{\partial(\partial_x j)}\right)}^0 + \partial_x f = \frac{\partial \mathcal{L}}{\partial j}. \qquad (28)$$

Equations (25) and (26) above yield the two constraints, the balance equation (16) and the value of the integrated dissipation (18), so we do not repeat them here. Besides, and again in order not to clutter our expressions, note that we drop hereafter the subindex 0 in the solutions of the variational problem, as is usual in physics. The partial derivatives of $\mathcal{L}$ that appear in the Euler-Lagrange equations are just

$$\frac{\partial \mathcal{L}}{\partial(\partial_x \rho)} = D(\rho)\frac{j + D(\rho)\partial_x \rho}{\sigma(\rho)}, \quad \frac{\partial \mathcal{L}}{\partial \rho} = D'(\rho)\partial_x \rho \frac{j + D(\rho)\partial_x \rho}{\sigma(\rho)} - \sigma'(\rho)\frac{[j + D(\rho)\partial_x \rho]^2}{2\sigma^2(\rho)} \qquad (29)$$

$$\frac{\partial \mathcal{L}}{\partial j} = \frac{j + D(\rho)\partial_x \rho}{\sigma(\rho)}, \qquad (30)$$

Note that if both $f$ and $\lambda$ are identically zero, the resulting Euler-Lagrange equations lead to the average solution, for which

$$j + D(\rho)\partial_x \rho = 0, \qquad (31)$$

and all the derivatives of the Lagrangian in (29) and (30) vanish. Of course, for this unconstrained variational problem the integrated dissipation corresponds to the average value. On the other hand, a separation from the average behaviour implies that the above Lagrangian derivatives do not vanish and, in general, both $f$ and $\lambda$ are different from zero.

### 4.1 Boundary conditions

It is clear that a dissipative system, in the absence of some external mechanism that injects energy into the system, will eventually reach in the long-time limit a trivial state of *thermal death* due to the continuous dissipation of energy in the bulk. On the other hand, if an energy-injection mechanism or *thermostat* is present, the system of interest may reach a non-equilibrium steady state in the long time limit in which the external energy input and the bulk energy loss balance each other.

There are many different types of thermostats, like e.g. stochastic bulk thermostats that homogeneously inject energy to all particles in the system via a random Langevin force, but also boundary thermostats as e.g. vibrated walls in granular media [11–19]. In this paper, and to be specific, we consider a dissipative system thermostatted at the boundaries. At the mesoscopic level this translates into an energy field which is fixed at the boundaries at any time,

$$\rho(x = \pm 1/2, t) = T \qquad \forall t \in [0, \tau]. \qquad (32)$$

For the sake of simplicity, we are considering that the temperature $T$ at both ends is the same. Even in this case, the system behaviour is highly non-trivial because the boundary character of the energy injection allows the system to develop gradients in its bulk, controlled by the local dissipation rate[3]. In fact, in the long time limit, the system would reach a non-trivial steady state characterised by an inhomogeneous density profile $\rho^{\rm st}(x)$ solution of the following differential equation,

$$\frac{d}{dx}\left[D(\rho^{\rm st})\frac{d}{dx}\rho^{\rm st}(x)\right] = \nu R(\rho^{\rm st}), \tag{33}$$

subject to the boundary conditions (32).

The question we are interested in now concerns the suitable boundary conditions for the Euler-Lagrange equations derived above, under the boundary injection mechanism just described. The simplest variational problems would appear if either (i) the unknown functions had fixed values at the boundary of the region of interest (in our case, the rectangle $[0, \tau] \times [-1/2, 1/2]$ in the $t - x$ plane) or (ii) one had periodic boundary conditions. In both cases, the boundary terms that appear when a variation $\delta\mathcal{S}$ is considered would vanish, in case (i) because the variations of the unknown functions would be zero at the boundaries and in case (ii) because the two boundary terms would cancel each other. Therefore, the boundary conditions for the Euler-Lagrange equations would be of Dirichlet type in case (i) or periodic in case (ii).

When the values of the unknown functions are neither fixed nor periodically repeated at the boundaries, the extremum condition $\delta\mathcal{S} = 0$ provides itself the right boundary conditions [20], as the terms multiplying the variations of the unknown function at the boundaries must be zero. In the thermostatted system, the values of the energy density are fixed at the boundaries $x = \pm 1/2$, but the values of the current are not, so the terms that multiply $\delta j(x = \pm 1/2)$ in the variation $\delta\mathcal{S}$ must vanish. These terms define the *momentum* $\tilde{p}_j$ conjugate to the current for the constrained Lagrangian $\tilde{\mathcal{L}}$,

$$\tilde{p}_j = \frac{\partial\tilde{\mathcal{L}}}{\partial(\partial_x j)} = \frac{\partial\mathcal{L}}{\partial(\partial_x j)} + f, \tag{34}$$

so that

$$\tilde{p}_j(x = \pm 1/2) = 0. \tag{35}$$

The variational calculation sketched here defines a complex spatiotemporal problem whose solution remains challenging in the general case. We hence need additional simplifying hypotheses. In particular, we will now assume that the optimal fields associated to a given fluctuation are in fact time-independent. This hypothesis is the generalisation of the *additivity principle* of Bodineau and Derrida [21] to the problem of fluctuations in driven dissipative media. Interestingly, the validity of this conjecture has been confirmed for a wide range of fluctuations, both in conservative systems [22,23] and dissipative media [7,8]. [4] Using now the additivity conjecture, and in particular the time-independence of optimal fields, it is easy to show that the balance equation reduces to

$$\frac{d}{dx}j(x) + \nu R(\rho(x)) = 0, \tag{36}$$

---

[3] This is not the case for bulk thermostats, where energy is injected homogeneously and thus the system remains spatially homogeneous.

[4] Nevertheless, this additivity scenario may break down for very large fluctuations in isolated systems, a regime in which a dynamic phase transition accompanied by a spontaneous symmetry breaking phenomenon has been predicted and observed under conservative dynamics [7,24–27].

In this way, the fixed value of the dissipation is just

$$d = \nu \int_{-1/2}^{1/2} dx\, R(\rho(x)) = j_l - j_r, \qquad \text{with } j_{r,l} \equiv j(x = \pm 1/2). \tag{37}$$

Moreover, the symmetry of the problem around $x = 0$ suggests that the optimal profiles are of definite parity: since $\rho(x = \pm 1/2) = T$, we look for solutions of the variational problem in which $\rho$ is an even function of $x$ and hence $j$ is and odd function. Therefore, $j_r = -j_l$ in (37), which leads to

$$j_l = d/2, \quad j_r = -d/2. \tag{38}$$

Interestingly, the previous arguments show that the constraint given by the fixed value of the integrated dissipation (14) can be mapped onto a specific set of boundary conditions for the current field. Moreover, the remaining constraint, that is, the stationary balance equation (36), can be used to eliminate the field $\rho$ in favour of the current $j$. This leads to a simpler unconstrained variational problem with this specific set of boundary conditions, instead of the more complex constrained variational problem described in this section.

## 5 The unconstrained variational problem with a second-order-derivative Lagrangian

In order to eliminate the density and write down a closed variational problem for the current, we first need some definitions. We introduce

$$y = R(\rho), \tag{39}$$

such that

$$\mathrm{d}(x,t) = -\nu y(x,t). \tag{40}$$

In terms of the new field $y$, we can write the balance equation as

$$j'(x) + \nu y(x) = 0, \tag{41}$$

with $'$ denoting first derivative with respect to the argument. Fourier's law for the average current just reads

$$j^{\mathrm{st}}(x) = -\hat{D}(y^{\mathrm{st}}(x)) \frac{d}{dx} y^{\mathrm{st}}(x), \tag{42}$$

with

$$\hat{D}(y) = \left(\frac{dy}{d\rho}\right)^{-1} D(\rho), \tag{43}$$

and $y^{\mathrm{st}}(x) \equiv R(\rho^{\mathrm{st}})$. Note that, according to the additivity principle introduced earlier, we are dropping the possible time dependence of all the fields.

Interestingly, we may now eliminate the density field in favour of the current by using Eq. (41), i.e. $y(x) = -j'(x)/\nu$. With this substitution, the rate function for the dissipation in (20) can be written as

$$G(d) = -\min_{j(x)} \mathcal{S}(j), \quad \text{with} \quad \mathcal{S}(j) = \int_{-1/2}^{1/2} dx\, \mathcal{L}(j, j', j''), \tag{44}$$

with a generalised Lagrangian

$$\mathcal{L}(j, j', j'') = \frac{\left[j - \hat{D}(-j'/\nu)\frac{j''}{\nu}\right]^2}{2\,\hat{\sigma}(-j'/\nu)}.$$ (45)

In the above expressions, $\hat{D}$ is the effective diffusivity, just defined in eq. (43), and $\hat{\sigma}$ is the mobility, introduced in (5), but written in terms of $y = -j'/\nu$. The generalised Lagrangian $\mathcal{L}(j, j', j'')$ depends both on the first and second derivatives of the current. Since the constraint of a given value of the integrated dissipation may be mapped onto certain boundary condition for the current, see (38), we seek the solution of the following generalised Euler-Lagrange equation [39]

$$\frac{d^2}{dx^2}\left(\frac{\partial \mathcal{L}}{\partial j''}\right) - \frac{d}{dx}\left(\frac{\partial \mathcal{L}}{\partial j'}\right) + \frac{\partial \mathcal{L}}{\partial j} = 0.$$ (46)

The boundary conditions come from the thermostatted boundaries and the imposed integrated dissipation. Taking into account (41), (32) and (38),

$$j'(\pm 1/2) = -\nu R(T), \quad j(-1/2) = -j(1/2) = d/2.$$ (47)

The equivalence of this simplified variational problem with the more general scheme of Section 4 can be established rigorously, for details see [8].

### 5.1 Equivalent Hamiltonian description

It is possible to define a Hamiltonian from a Lagrangian with second-order derivatives [20]. This Hamiltonian description allows us to write down a set of four first-order differential equations, which of course lead to the same solution of the variational problem as the fourth order Euler-Lagrange equation (46), but which in some cases turn out to simplify its solution. Here we sketch the procedure, further details can be found in either [20] or [8]. The canonical coordinates are $y$ and $j$, and the suitable conjugate canonical momenta are

$$p_y \equiv -\nu \frac{\partial \mathcal{L}}{\partial j''}, \quad p_j = \frac{\partial \mathcal{L}}{\partial j'} - \frac{d}{dx}\left(\frac{\partial \mathcal{L}}{\partial j''}\right).$$ (48)

The Euler-Lagrange equation may then be written as $dp_j/dx = \partial \mathcal{L}/\partial j$, formally identical to the usual case. The Hamiltonian follows as

$$\mathcal{H} = y p_y + j p_j - \mathcal{L},$$ (49)

and in our particular situation, by using (19), which can be rewritten as $\mathcal{L} = Q(y)p_y^2/2$, and (48), we obtain

$$\mathcal{H} = \frac{1}{2}Q(y)p_y^2 - \hat{D}^{-1}(y)jp_y - \nu y p_j,$$ (50)

$$Q(y) \equiv \frac{\hat{\sigma}(y)}{\hat{D}^2(y)}.$$ (51)

The optimal profiles that sustain the integrated dissipation fluctuation obey the canonical equations,

$$y' = \frac{\partial \mathcal{H}}{\partial p_y} = Q(y)p_y - \hat{D}^{-1}(y)j,$$ (52)

$$j' = \frac{\partial \mathcal{H}}{\partial p_j} = -\nu y, \tag{53}$$

$$p'_y = -\frac{\partial \mathcal{H}}{\partial y} = -\frac{dQ(y)}{dy}\frac{p_y^2}{2} + \frac{d\hat{D}^{-1}(y)}{dy}jp_y + \nu p_j, \tag{54}$$

$$p'_j = -\frac{\partial \mathcal{H}}{\partial \hat{j}} = \hat{D}^{-1}(y)p_y, \tag{55}$$

with boundary conditions

$$y(\pm 1/2) = R(T), \quad j(-1/2) = -j(1/2) = d/2. \tag{56}$$

The rate function (44) may be cast in the following form,

$$G(d) = -\int_{-1/2}^{1/2} dx\,\mathcal{L} = -\frac{1}{2}\int_{-1/2}^{1/2} dx\, Q(y)p_y^2, \tag{57}$$

in which $y$ and $p_y$ have to be evaluated over the solutions of the canonical equations for the corresponding value of $d$.

It should be emphasised how the average behaviour is obtained in the canonical description. If the values of the variables are not fixed at the boundaries, the associated conjugate canonical momenta must vanish there (since the boundary terms in the variation $\delta\mathcal{S}$ have the form of the variation of the coordinate multiplied by the conjugate canonical momentum [20]). Then, it is natural that that the canonical equations have a particular solution with $p_y = 0$ and $p_j = 0$ for all $x$: this solution leads to $G(d) = 0$, which corresponds to the average behaviour. By substituting $p_y = 0$ and $p_j = 0$ into (52)-(55), we get that $y' = -j/\hat{D}(y)$, $j' = -\nu y$. These two equations are just equivalent to (33) for the average profiles in the steady state.

# 6 Solving the problem: analytics, numerical strategies and simulation of rare events

## 6.1 Analytical solution in a limiting case

In order to give a flavour of the physics behind the formalism here presented, we now sketch the solution of the MFT problem for the fluctuations of the empirical dissipation in the limit of weak dissipation, $\nu \ll 1$, for the linear dissipative KMP model described in Section 2. Recall that this particular model is characterised by a constant diffusivity $D(\rho) = 1/2$, a mobility $\sigma(\rho) = \rho^2$ and a dissipation transport coefficient $R(\rho) = \rho$. The steady-state, average solution to Eqs. (33) and (36) can be simply derived in this case, and reads

$$\rho^{\text{st}}(x) = T\frac{\cosh(\sqrt{2\nu}x)}{\cosh(\sqrt{\nu/2})}, \qquad j^{\text{st}}(x) = -T\sqrt{\frac{\nu}{2}}\frac{\sinh(\sqrt{2\nu}x)}{\cosh(\sqrt{\nu/2})}, \tag{58}$$

Moreover, the average dissipation is $\langle d \rangle \equiv \nu \int_{-1/2}^{1/2} dx\rho^{\text{st}}(x) = T\sqrt{2\nu}\tanh(\sqrt{\nu/2})$. It is interesting to note that these steady-state density and current profiles are similar those found in vibrated granular gases, when space is measured in units of the mean-free-path [28].

Our aim is to obtain analytically, from the MFT formalism described above, the whole dissipation large deviation function for this model in the weakly-dissipative limit $\nu \ll 1$. For that, we first particularise the general equations of Section 5 for this specific problem. In particular, Eq. (46) results in the following differential equation for the optimal current field

$$12\nu^2 j(x)(j'^2(x) - j(x)j''(x)) + j''(x)(3j''(x) - 4j'(x)j'''(x)) + j'^2(x)j''''(x) = 0 \,. \quad (59)$$

with boundary conditions following from Eqs. (32) and (38),

$$j'(\pm 1/2) = -\nu R(T), \quad j(-1/2) = -j(1/2) = d/2 \,. \quad (60)$$

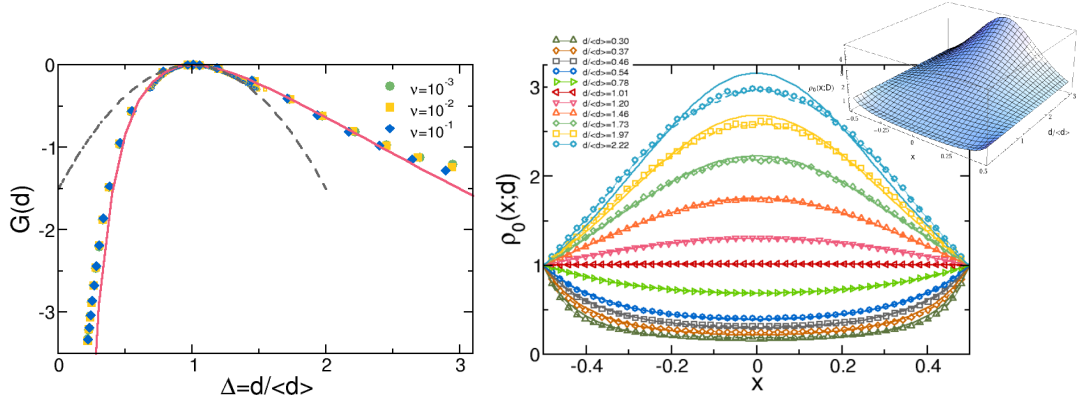

**Fig. 2.** (Color online) Left: Scaling of the dissipation LDF in the quasi-elastic limit ($\nu \ll 1$) for $N = 50$, $T = 1$ and varying $\nu$. The solid and dashed lines are the MFT prediction and Gaussian estimation, respectively. Small points around the peak were obtained in standard simulations, while the bigger points come from advanced cloning simulations, see Section 6.3 below. Right: Optimal energy profiles for varying $d/\langle d \rangle$ measured for $\nu = 10^{-3}$ (symbols) and $\nu = 10^{-2}$ (dashed lines), and MFT predictions (solid lines and inset).

The behaviour of the steady-state current field and average dissipation for $\nu \ll 1$, see Eq. (58) and paragraph below, strongly suggests the scaling form $j(x) = \nu\psi(x)$ and $d = \nu\Delta$ in this limit, respectively, where $\psi(x)$ and $\Delta$ (the total energy per site) remain of order one even though $\nu \ll 1$. In fact, $\langle d \rangle \sim \nu T$ for $\nu \ll 1$, i.e. $\Delta \sim T$. Using this scaling in Eq. (59) and retaining only the lowest order in $\nu$ we get a linear differential equation for $\psi(x)$ with solution

$$\psi_0(x; d) = -\frac{T}{b}\frac{\tanh(bx)}{1 - \tanh^2\left(\frac{b}{2}\right)}, \quad \rho_0(x; d) = T\frac{\cosh^2\left(\frac{b}{2}\right)}{\cosh^2(bx)}, \quad (61)$$

where $\rho_0(x; d) = -\psi_0'(x; d)$, and $b$ is a constant implicitly given by the following transcendental equation

$$\frac{\sinh b}{b} = \frac{d}{\langle d \rangle} \,. \quad (62)$$

The average behaviour in this weakly-dissipative regime is recovered in the limit $b \to 0$ (i.e. $d \to \langle d \rangle$). For $b \ll 1$ we have small deviations from the average and Gaussian statistics as expected. For arbitrary values of $b$, eq. (61) gives the optimal profiles for the system to sustain an arbitrary fluctuation of the dissipated energy $d$. Using these optimal profiles in Eqs. (44)-(45) we find for $\nu \ll 1$

$$G(d) = b\tanh\left(\frac{b}{2}\right) - \frac{b^2}{2}, \quad (63)$$

with $b$ implicitly given by eq. (62) in terms of $d/\langle d \rangle$. This means that, in the weak dissipation limit $\nu \ll 1$, $G(d)$ shows a simple scaling form independent of $\nu$ when plotted against the *relative* dissipation $d/\langle d \rangle$. This scaling form is fully confirmed in Fig. 2, left panel, which shows $G(d)$ as measured for different, small values of $\nu \in [10^{-3}, 10^{-1}]$ in advanced cloning Monte Carlo simulations of the linear dissipative KMP model, see Section 6.3 below for simulation details. Interestingly, the dissipation LDF is highly skewed with a sharp decrease for fluctuations $d < \langle d \rangle$ and no negative branch, so relations similar to the fluctuation theorem linking the probabilities of a given dissipation $d$ and the inverse event $-d$ do not hold [29,30]. This was of course expected from the lack of micro-reversibility in the microscopic dissipative model, a basic tenet for the fluctuation theorem to apply [31]. The limit $b \gg 1$ corresponds to large dissipation fluctuations, where $G(d) \approx -\frac{1}{2}[\ln(d/\langle d \rangle)]^2$, i.e. a very slow decay which shows that such large fluctuations are far more probable than expected within Gaussian statistics ($\sim -\frac{3}{2}(d/\langle d \rangle)^2$). We also measured the optimal energy density profile associated to a given dissipation fluctuation, see right panel in Fig. 2, finding also very good agreement with MFT predictions. Note that optimal profiles for varying $\nu \ll 1$ also collapse for constant $d/\langle d \rangle$. Moreover, profiles associated to dissipation fluctuations above the average exhibit an energy overshoot in the bulk. This observation suggests that the mechanism responsible for large dissipation fluctuations consists in a continued over-injection of energy from the boundary bath, which is transported to and stored in the bulk before being dissipated.

For not-so-weak values of the dissipation coefficient, $\nu \gtrsim 1$, the analytical solution of the non-linear differential equation (59) for the optimal current profile is unfeasible, so we have to turn to numerical schemes of solution. This technique is briefly described now for the ease of the reader, and we finish this Section with a detailed exposition of advanced methods to measure rare events in Monte Carlo simulations of lattice gases.

## 6.2 Numerical strategy: the shooting method

As stated above, for $\nu \gtrsim 1$ the perturbative method of the previous Section does not work, and we have to solve the fully non-linear differential equation for the optimal current profile. The strong non-linearity prevents any analytical solution to the problem in general, so we have to turn to numerical schemes of solution. Here we present in a nutshell a standard numerical method as applied to solve the fourth-order differential equation (59) for the linear dissipative KMP model, with boundary conditions (60).

To address this problem we use the *shooting method* for a set of $N$ coupled first-order differential equations. In our particular case, the first step consists in converting our $4^{\text{th}}$-order non-linear differential equation for $j(x)$, see Eq. (59), into 4 coupled first-order differential equation. For that, we define three new functions $k(x) \equiv j'(x)$, $l(x) \equiv j''(x)$ and $m(x) \equiv j'''(x)$, so that the new system of first-order differential equations trivially reads

$$12\nu^2 j(x)\left[k^2(x) - j(x)l(x)\right] + l(x)\left[3l(x) - 4k(x)m(x)\right] + k^2(x)m'(x) = 0\,,$$
$$j'(x) = k(x)\,, \qquad\qquad k'(x) = l(x)\,, \qquad\qquad l'(x) = m(x).$$

In order to integrate this set of differential equations in the interval $x \in [-1/2, 1/2]$, we must provide four initial conditions. Instead, we have *boundary conditions* for both $j(x)$ and $j'(x) \equiv k(x)$, see Eq. (60). The shooting method consists in trading the two boundary conditions for $j(x)$ and $k(x)$ at $x = +1/2$ for two initial conditions for $l(x)$ and $m(x)$ at $x = -1/2$. As the relation between these two sets of conditions is not known a priori, a recursive procedure is designed to estimate the initial conditions $l_- = l(x = -1/2)$ and $m_- = m(x = -1/2)$ for $l(x)$ and $m(x)$ leading to the correct boundary conditions for $j(x)$ and $k(x)$ at $x = +1/2$. For that, we define two functions $F_1(l_-, m_-) = j(x = 1/2) + d/2$ and $F_2(l_-, m_-) = k(x = 1/2) + \nu R(T)$, which measure the distance of the actual boundary values of the fields $l(x)$ and $m(x)$ at $x = 1/2$ from the desired values, see Eq. (60), and find their root using a recursive strategy, based on e.g. the Newton-Raphson algorithm [32], to improve at each step the estimated initial conditions. A good starting point for this recursive algorithm is given by the initial values associated to the steady-state average solution for the current field,

$$ l_- = j''_{st}(x = -1/2), \qquad m_- = j'''_{st}(x = -1/2), \tag{64} $$

with $j^{st}(x)$ given in Eq. (58). This process leads to a reliable numerical estimation of the optimal current (and density) profile for a given $d$, which we can use together with Eqs. (44)-(45) to estimate the value of $G(d)$.

Appendix 7 summarises a Mathematica® notebook to solve completely this numerical problem, obtaining along the way the dissipation LFD in the non-perturbative case $\nu \gtrsim 1$.

### 6.3 Simulation of rare events

In general, large deviation functions are very hard to measure in experiments or simulations because they involve by definition exponentially-unlikely events, i.e. $P_t(d) \sim \exp[+tLG(d)]$. In this way, observing a significant long-time fluctuation away from the average dissipation would need an exceedingly large number of $N \sim e^t$ simulations to gather enough statistics to actually observe such rare event. This problem can be solved at least partially by implementing a general strategy dubbed *"Go with the winners"* [33], based on (i) the modification of the underlying dynamics so the rare event in the standard dynamics becomes a typical event in the modified process, together with (ii) a clone population dynamics favouring particular system copies with the desired behaviour. The algorithm here described, as applied to the numerical evaluation of large deviation functions, was first proposed by Kurchan and coworkers in [34], and has been expanded to continuous-time Markov systems in [35]; see [36] for a recent review, including applications to deterministic systems.

We now explain the method as applied to study typical and rare fluctuations of the space- and time-integrated dissipation in the linear dissipative KMP model described in Section 2. Let $U_{C'C}$ be the transition rate from configuration $C$ to $C'$ for our model, with $C = \{\rho_i, i \in [1, L]\}$ and $\rho_i$ the energy at site $i$ in a 1D lattice of linear size $L$. The probability of measuring an integrated dissipation $D_t$ after a total time $t$ starting from a configuration $C_0$ can be written as

$$ P(D_t, t; C_0) = \sum_{C_t \cdots C_1} U_{C_t C_{t-1}} \cdots U_{C_1 C_0} \delta(D_t - \sum_{k=0}^{t-1} D_{C_{k-1} C_k}), \tag{65} $$

where $D_{C'C}$ is the elementary dissipation involved in the microscopic transition $C \to C'$. For our particular model, this elementary dissipation in a collision between sites $(i, i + 1)$ equals $-(1 - \alpha)(\rho_i + \rho_{i+1})/(L - 1)$, with $\alpha$ the restitution coefficient, see Eq. (8). For long times we expect the information on the initial state $C_0$ be lost, so $P(D_t, t; C_0) \to P(D_t, t)$. In this limit $P(D_t, t)$ obeys a large deviation principle $P(D_t, t) \sim \exp[+tLG(d = D_t/t)]$.

Note however that, in most cases, it is convenient to work with the moment generating function of the above distribution, defined as

$$\Pi(\lambda, t) = \sum_{D_t} e^{\lambda D_t} P(D_t, t) = \sum_{C_t \cdots C_1} U_{C_t C_{t-1}} \cdots U_{C_1 C_0} e^{\lambda \sum_{k=0}^{t-1} D_{C_{k-1} C_k}}, \qquad (66)$$

where we have applied Eq. (65) in the second equality. For long $t$, it is easy to show that $\Pi(\lambda, t) \to \exp[+t\mu(\lambda)]$, with $\mu(\lambda) = \max_d[G(d) + \lambda d]$. The previous equation suggest to define a modified dynamics $\tilde{U}_{C'C}(\lambda) \equiv e^{\lambda D_{C'C}} U_{C'C}$, and therefore

$$\Pi(\lambda, t) = \sum_{C_t \cdots C_1} \tilde{U}_{C_t C_{t-1}} \cdots \tilde{U}_{C_1 C_0}. \qquad (67)$$

It is important to stress that this new dynamics is unnormalised, $\sum_{C'} \tilde{U}_{C'C} \neq 1$.

It is useful to introduce at this point Dirac's bra and ket notation, together with the quantum Hamiltonian formalism for the master equation [37,38]. The idea is to assign to each system configuration $C$ a vector $|C\rangle$ in configuration space which, together with its transposed vector $\langle C|$, form an orthogonal basis of a complex space and its dual [37,38]. For example, in the simplest case with a finite number of available configurations (not our case), one could write $|C\rangle^T = \langle C| = (\cdots 0 \cdots 0, 1, 0 \cdots 0 \cdots)$, i.e. all components equal to zero except for the component corresponding to configuration $C$, which is 1. In this notation, $\tilde{U}_{C'C} = \langle C'|\tilde{U}|C\rangle$ and the probability distribution is written as a probability vector

$$|P(t)\rangle = \sum_C P(C, t)|C\rangle, \qquad (68)$$

where $P(C, t) = \langle C|P(t)\rangle$, with the scalar product $\langle C'|C\rangle = \delta_{C'C}$. If $\langle s| = (1\cdots 1)$, normalisation then implies $\langle s|P(t)\rangle = 1$.

With the previous notation, we can now write the spectral decomposition of operator $\tilde{U}(\lambda) = \sum_n e^{\Lambda_n(\lambda)} |\Lambda_n^R(\lambda)\rangle\langle\Lambda_n^L(\lambda)|$ where we assume that a complete biorthogonal basis of right and left eigenvectors for matrix $\tilde{U}$ exists, $\tilde{U}|\Lambda_n^R(\lambda)\rangle = e^{\Lambda_n(\lambda)}|\Lambda_n^R(\lambda)\rangle$ and $\langle\Lambda_n^L(\lambda)|\tilde{U} = e^{\Lambda_n(\lambda)}\langle\Lambda_n^L(\lambda)|$. Denoting as $e^{\Lambda(\lambda)}$ the largest eigenvalue of $\tilde{U}(\lambda)$, with associated right and left eigenvectors $|\Lambda^R(\lambda)\rangle$ and $\langle\Lambda^L(\lambda)|$ respectively, and writing $\Pi(\lambda, t) = \sum_{C_t} \langle C_t|\tilde{U}^t|C_0\rangle$, we find for long times

$$\Pi(\lambda, t) \to e^{+t\Lambda(\lambda)} \langle\Lambda^L(\lambda)|C_0\rangle \left( \sum_{C_t} \langle C_t|\Lambda^R(\lambda)\rangle \right). \qquad (69)$$

In this way we have $\mu(\lambda) = \Lambda(\lambda)$, so the Legendre transform of the dissipation LDF is given by the natural logarithm of the largest eigenvalue of $\tilde{U}(\lambda)$. In order to measure this eigenvalue in Monte Carlo simulations, and given that the dynamics $\tilde{U}$ is unnormalised, we introduce now the exit rates $Y_C = \sum_{C'} \tilde{U}_{C'C}$ and define the normalised dynamics $U'_{C'C} \equiv Y_C^{-1} \tilde{U}_{C'C}$. In this way

$$\Pi(\lambda, t) = \sum_{C_t \cdots C_1} Y_{C_{t-1}} U'_{C_t C_{t-1}} \cdots Y_{C_0} U'_{C_1 C_0}. \qquad (70)$$

This sum over paths can be realised by considering an ensemble of $M \gg 1$ copies or clones of the system, evolving sequentially according to the following Monte Carlo scheme:

1. Each copy evolve independently according to modified normalised dynamics $U'_{C'C}$.

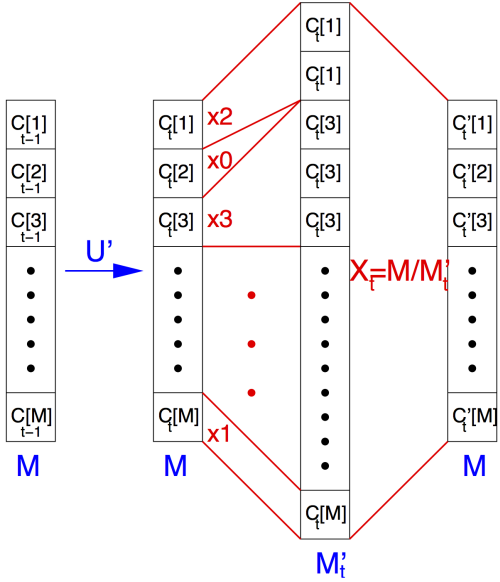

**Fig. 3.** Sketch of the evolution and cloning of the copies during the evaluation of the large deviation function.

2. Each copy $m \in [1, M]$ (in configuration $C_t[m]$ at time $t$) is cloned with rate $Y_{C_t[m]}$. This means that, for each copy $m \in [1, M]$, we generate a number $K_{C_t[m]} = \lfloor Y_{C_t[m]} \rfloor + 1$ of identical clones with probability $Y_{C_t[m]} - \lfloor Y_{C_t[m]} \rfloor$, or $K_{C_t[m]} = \lfloor Y_{C_t[m]} \rfloor$ otherwise, where $\lfloor x \rfloor$ represents the integer part of $x$. Note that if $K_{C_t[m]} = 0$ the copy may be killed and leave no offspring. This procedure gives rise to a total of $M_t' = \sum_{m=1}^{M} K_{C_t[m]}$ copies after cloning all of the original $M$ copies.
3. Once all copies evolve and clone, the total number of copies $M_t'$ is sent back to $M$ by an uniform cloning probability $X_t = M/M_t'$.

Fig. 3 sketches this procedure. It can be shown that, for long times, we recover $\mu(\lambda)$ via

$$\mu(\lambda) = -\frac{1}{t} \ln(X_t \cdots X_0), \qquad t \gg 1. \tag{71}$$

In order to derive this expression, first consider the cloning dynamics above, but without keeping the total number of clones constant (that is, forgetting about step 3). In this case, for a given history $\{C_t, \cdots, C_0\}$, the number $N(C_t \cdots C_0, t)$ of clones in configuration $C_t$ at time $t$ will be $N(C_t \cdots C_0, t) = Y_{C_{t-1}} U'_{C_t C_{t-1}} N(C_{t-1} \cdots C_0, t-1)$, so that

$$N(C_t \cdots C_0, t) = Y_{C_{t-1}} U'_{C_t C_{t-1}} \cdots Y_{C_0} U'_{C_1 C_0} N(C_0, 0). \tag{72}$$

Summing over all histories of duration $t$, see Eq. (70), we find that the average of the total number of clones at long times shows exponential behaviour, $\langle N(t) \rangle = \sum_{C_t \cdots C_1} N(C_t \cdot C_0, t) \sim N(C_0, 0) exp[+t\mu(\lambda)]$. Now, going back to step 3, when the fixed number of copies $M$ is large enough, we have $X_t = \langle N(t-1) \rangle / \langle N(t) \rangle$ for the global cloning factors, so $X_t \cdots X_1 = \langle N(C_0, 0) \rangle / \langle N(t) \rangle$ and we recover expression (71) for $\mu(\lambda)$.

We now particularise this method to measure the dissipated energy LDF for the linear dissipative KMP model of Section 2. In this case, the transition rate from a configuration $C = \{\rho_1 \cdots \rho_N\}$ to another configuration $C'_y = \{\rho_1 \cdots \rho'_y, \rho'_{y+1} \cdots \rho_N\}$, with $y \in [0, L]$ where the pair $(\rho'_y, \rho'_{y+1})$ has collided following the rules of Eq. (9),

can be written as

$$U_{C'_y C} = \begin{cases} (L+1)^{-1}, \ y \in [1, L-1] \\ \frac{\beta e^{\beta \rho_1}}{L+1} E_1[\beta \max(\rho_1, \rho'_1)], \ y = 0 \\ \frac{\beta e^{\beta \rho_L}}{L+1} E_1[\beta \max(\rho_L, \rho'_L)], \ y = L \end{cases} \tag{73}$$

Here $E_1(x) = -Ei(-x)$, were $Ei(x)$ is the exponential integral function, or

$$E_1(x) = \int_x^\infty du \frac{e^{-u}}{u} \tag{74}$$

This appears when integrating over all possible pairs $(z, \tilde{\rho}_{L,R})$ that can result in a given $\rho'_{1,L}$ respectively, see Eq. (10). It is straightforward to show that $U_{C'_y C}$ is normalised. In order to measure the dissipated energy fluctuations we recall the definition of the dissipation produced at each microscopic step. Taking in account Eq. 8 , we can write

$$D_{C'_y C} = \begin{cases} \frac{-(1-\alpha)(\rho_y + \rho_{y+1})}{L-1}, \ y \in [1, L-1] \\ 0, \qquad y = 0, L \end{cases} \tag{75}$$

In this way, we can measure the dissipated energy in the system. Using this definition we can write the normalised modified dynamics as $U'_{C'_y C} \equiv Y_C^{-1} U_{C'_y C} \, e^{\lambda D_{C'_y C}}$, and for $y \in [1, L-1]$ we have

$$U'_{C'_y C} = \frac{e^{-\overline{\lambda}(1-\alpha)(\rho_y + \rho_{y+1})}}{Y_C(L+1)} \tag{76}$$

with $\overline{\lambda} = \lambda/(L-1)$, while for $y = 0, L$ we obtain $U'_{C'_y C} \equiv Y_C^{-1} U_{C'_y C}$. Finally, the exit rate is given by

$$Y_C = \frac{2}{L+2} + \sum_{y=1}^{L-1} \frac{e^{-\overline{\lambda}(1-\alpha)(\rho_y + \rho_{y+1})}}{L+1} \tag{77}$$

Using this method, we have measured the dissipation LDF for the 1d linear dissipative KMP model shown in Fig. 2, as well as the optimal density profiles associated to different dissipation fluctuations (which are related with both the right and left eigenvectors linked to the leading eigenvalue of the modified dynamics $\tilde{U}$, see ref. [27] for further details.

## 7 Conclusions

In these notes, we have attempted to present a brief pedagogical account of the fluctuation behaviour of driven dissipative systems, as understood from a suitable generalisation of macroscopic fluctuation theory. The results here described possess a general character, as they apply to many different non-linear reaction-diffusion systems where dissipation and diffusion compete at mesoscopic space and time scales, as captured by the balance equation (1). The Euler-Lagrange equations that stem from the general MFT variational problem are complex, and need as input the transport coefficients for the problem at hand. The latter can be obtained by means of coarse graining procedures for microscopic stochastic models, or on phenomenological or experimental grounds. On the other hand, it is now feasible to carry out numerical experiments to measure arbitrary fluctuations; recent advanced cloning Monte Carlo techniques make it possible to simulate rare events in an efficient way [34].

Analytical results can be obtained by following the general scheme developed above in some cases. More specifically, for a general class of lattice models, the balance equation and the transport coefficients can be derived from the microscopic dynamics [6], and a substantial study of the statistics of dissipation fluctuations, including rare event simulation and analytical results, can be found in [8]. In particular, the variational problem can be solved analytically in some different physical situations of interest, for example in the limit of weakly dissipative systems, as described in previous sections.

The results reported here open new interesting possibilities for future research. To mention just a few, it would be interesting to investigate dissipation fluctuations under different boundary conditions, or to understand fluctuations of other quantities different from the dissipated energy. To illustrate these points, it seems worth exploring dissipative systems thermostatted at only one of their ends, including the semi-infinite case, as for these systems the symmetry around the origin is broken. In addition, current fluctuations and joint current-dissipation fluctuations in driven dissipative media deserve further investigation.

An interesting case, not discussed here, is the possibility of $\nu < 0$ in Eq. (1). This negative dissipation coefficient would mimic a local energy source mechanism, much in the spirit of *self-propulsion* in active matter. It would be interesting to study in detail the implications of the formalism for fluctuations here developed in the $\nu < 0$, *active-matter* limit. This is left for future work.

Finally, it is worth mentioning that the generalisation of macroscopic fluctuation theory to driven dissipative media, as put forward here, is limited to systems with a single, non-conserved hydrodynamic field. In real dissipative systems, as e.g. granular media, there are several coupled hydrodynamic fields, some of them are conserved (such as mass or momentum) whereas others are not (like the energy). It would be desirable to extend MFT to fluctuating granular hydrodynamics. In this regard, the generalisation of our theoretical framework to model systems that are simple but need more than one hydrodynamic field for their description, such as those in [10], would help to pave the way toward this goal.

## Acknowledgements

P. I. H. acknowledges financial support from Spanish project FIS2013-43201-P (MINECO), University of Granada, Junta de Andalucía project P09-FQM4682 and GENIL PYR-2014-13. A. P acknowledges the support of the Spanish Ministerio de Economía y Competitividad through grant FIS2014-53808-P.

## Appendix: Numerical evaluation with Mathematica®

In this appendix, we present an example of the numerical solution of the statistics of the dissipated energy, which is developed with Mathematica®. Specifically, we carry out the numerical integration of the boundary problem in a system with $\nu = 10$ and temperature at the boundaries $T_{l,r} = 1$. First, we introduce the Fourier law for the system of interest (see section 6),

```
In[1]:=  rho[x_] := (-Nu*j'[x]);
```

Second, the mobility

```
In[2]:=  Sigma[x] := (-(j'[x]/Nu))^2;
```

Third, the Lagrangian

```
In[3]:=   L[x] := (j[x] - j''[x]/(2 Nu))^2/(2*Sigma[x]);
```
and proceed with the corresponding Euler-Lagrange equation,
```
In[4]:=   equa[x] := FullSimplify[D[L[x],j[x]]+
          D[D[D[L[x],j''[x]],x],x]-D[D[ L[x],j'[x]],x]];
```
```
In[5]:=   y = equa[x];
```
Also, we set the values of some parameters, namely the boundary conditions for the temperature
```
In[6]:=   T = 1.;
```
and the macroscopic dissipation coefficient
```
In[7]:=   Nu = 10;
```
where Nu stands for $\nu$.

   In order to run the numerical procedure, we also have to introduce the average profiles for the energy density,
```
In[8]:=   yav[x_] := T*Cosh[x Sqrt[2 Nu]]/Cosh[Sqrt[Nu/2]];
```
the energy current,
```
In[9]:=   jave[x_] := -1/2*D[yav[x],x];
```
and the average dissipation field,
```
In[10]:=  dav := Nu Integrate[yav[x],{x,-1/2,1/2}];
```
Finally, we open the files where we will save the resulting profiles
```
In[11]:=  OpenWrite["totalprofilenu10k10_7.dat"];
```
and the resulting large deviation function $G(d)$
```
In[12]:=  str = OpenWrite["gd10k10_7.dat"];
```
   To conclude, we present the loop where the shooting procedure is run. This loop allows us to obtain the profiles and the large deviation function $G(d)$ for fixed values of $\nu$ and $T$.
```
In[13]:=  Do[ equa[x] == 0; boundary = {j[-0.5] == -Re[dav]*i/2,
          j'[-0.5] == -Nu,  j[0] == 0, j''[0] == 0};
          sol = NDSolve[{equa[x] == 0, boundary}, j, x,
          Method -> {"Shooting","StartingInitialConditions" ->
          {j[-0.5] == jave[x] /. x -> -0.5,
          j'[-0.5] == -Nu,
          j''[-0.5] == D[D[jave[x],x],x]/. x -> -0.5,
          j'''[-0.5] == D[D[D[jave[x],x],x],x]/. x -> -0.5,}}];
          f[x_] := j[x] /. sol;
          g[x_] :=D[f[u],u] /. u -> x;
          h[x_] :=D[g[u],u] /. u -> x;
          v[x_] :=D[h[u],u] /. u -> x;
          m[x_] :=D[k[u],u] /. u -> x;
          b[x_] := (-g[u]/Nu)^2 /. u -> x;
          l[x_] := -(f[x] - 1/2*h[x]/Nu)^2/(2*b[x]);
          int = NIntegrate[l[x], {x, -0.5, 0.5}];
          tab = Table[ N[{i, FortranForm[x],
          FortranForm[Re[-g[x]]/Nu]}, {x, 0., 0.5, 0.01}];
          AppendTo[perfiles, tab];
          Write[prof, OutputForm[TableForm[tab,
          TableSpacing -> {0, 3}]]];
          Write[prof, "    "]; Write[str, i, Re[int]];  delta = (i) ,
          {i, 0.1, 7., 0.08}]
```

In this loop, we have taken advantage of the parity of the solutions in the interval $x \in [-1/2, 1/2]$ to solve the problem in half the interval, namely $x \in [-1/2, 0]$ ($j$ is an odd function of $x$ whereas $\rho$ is even). Note that $d$ fixes the boundary for the current for a given fluctuation, see Eq. (47), and we are changing the value of $d$ by introducing $i$ such that $d = i\langle d \rangle$. In this particular example, $i \in [0.1, 7]$ ant its increment between neighbouring values $\Delta i = 0.08$. The latter defines the grid over which the LDF $G(d)$ is obtained. Moreover, the shooting procedure is started with the initial conditions provided by the average solution of the problem. Finally, once the numerical solution for the profiles is obtained, we integrate it to get $G(d)$.

## References

1. L. Bertini, A. De Sole, D. Gabrielli, G. Jona-Lasinio and C. Landim, Phys. Rev. Lett. **87**, (2001) 040601; Phys. Rev. Lett. **94**, (2005) 030601; J. Stat. Mech. (2007) P07014; J. Stat. Phys. **135**, (2009) 857; Rev. Mod. Phys. 87, (2015) 593.
2. R. S. Ellis, *Entropy, Large Deviations and Statistical Mechanics* (Springer, New York 1985).
3. B. Derrida, J. Stat. Mech. (2007) P07023.
4. H. Touchette, Phys. Rep. **478**, (2009) 1.
5. T. Pöschel and S. Luding eds., *Granular Gases*, Lecture Notes in Physics vol. 564 (Springer-Verlag, Berlin 2001).
6. A. Prados, A. Lasanta and P. I. Hurtado, Phys. Rev. E **86**, (2012) 031134.
7. A. Prados, A. Lasanta, and P. I. Hurtado, Phys. Rev. Lett. **107**, (2011) 140601.
8. P. I. Hurtado, A. Lasanta and A. Prados, Phys. Rev. E **88**, (2013) 022110.
9. C. Kipnis, C. Marchioro and E. Presutti, J. Stat. Phys. **27**, (1982) 65.
10. A. Lasanta, A. Manacorda, A. Prados, and A. Puglisi, New J. Phys. **17** (2015) 083039.
11. J. Javier Brey, M. J. Ruiz-Montero, and F. Moreno, Phys. Rev. E **63**, (2001) 061305.
12. P. Visco, A. Puglisi, A. Barrat, F. van Wijland and E. Trizac, Eur. Phys. J. B **63**, (2006) 377.
13. D. R. M. Williams and F. C. MacKintosh, Phys. Rev. E **54**, (1996) R9.
14. T. P. C. van Noije and M. H. Ernst, Granular Matter **1**, (1996) 57.
15. T. P. C. van Noije, M. H. Ernst, E. Trizac, and I. Pagonabarraga, Phys. Rev. E **63**, (1999) 4326.
16. A. Puglisi, V. Loreto, U. Marini Bettolo Marconi, A. Petri, and A. Vulpiani, Phys. Rev. Lett. **81**, (1998) 3848.
17. M. I. García de Soria, P. Maynar, and E. Trizac, Phys. Rev. E **85**, (2012) 051301.
18. M. I. García de Soria, P. Maynar, and E. Trizac, Phys. Rev. E **87**, (2013) 022201.
19. A. Prados and E. Trizac, Phys. Rev. Lett. **112**, (2013) 198001.
20. C. Lanczos, *The Variational Principles of Mechanics* (Dover, New York 1986).
21. T. Bodineau and B. Derrida, Phys. Rev. Lett. **92**, (2004) 180601.
22. P. I. Hurtado and P. L. Garrido, Phys. Rev. Lett. **102**, (2009) 250601; Phys. Rev. E **81**, (2010) 041102.
23. P. I. Hurtado, C. Pérez-Espigares, J. J. del Pozo and P. L. Garrido, Proc. Natl. Acad. Sci. USA **108**, (2011) 7704.
24. T. Bodineau and B. Derrida, Phys. Rev. E **72**, (2005) 066110.
25. P.I. Hurtado and P.L. Garrido, Phys. Rev. Lett. **107**, (2011) 180601.
26. C. P. Espigares, P. L. Garrido and P. I. Hurtado, Phys. Rev. E **87**, (2013) 032115.
27. P. I. Hurtado, C. Pérez-Espigares, J. J. del Pozo and P. L. Garrido, J. Stat. Phys. **154**, (2014) 214.
28. J.J. Brey, M.J. Ruiz-Montero and F. Moreno, Phys. Rev. E **62**, (2000) 5339.
29. G. Gallavotti and E.G.D. Cohen, Phys. Rev. Lett. **74**, (1995) 2694.
30. J.L. Lebowitz and H. Spohn, J. Stat. Phys. **95**, (1999) 333.
31. A. Puglisi et al, Phys. Rev. Lett. **95**, (2005) 110202.
32. W.H. Press, S.A. Teukolsky, T. Vetterling, and B.P. Flannery, *Numerical Recipes: The Art of Scientific Computing*, 3rd Edition (Cambridge University Press, New York 2007).

33. P.Grassberger. Comp. Phys. Com. **147**, (2002) 64.
34. C. Giardinà, J. Kurchan and L. Peliti, Phys. Rev. Lett. **96**, (2006) 120603.
35. V. Lecomte and J. Tailleur, J. Stat. Mech. (2007) P03004; AIP Conf. Proc. **1091**, (2009) 212.
36. C. Giardin, J. Kurchan, V. Lecomte and J. Tailleur, J. Stat. Phys. **145**, (2011) 787.
37. H. J. Harris, G. M. Schütz, J. Stat. Mech. (2007) P07020.
38. G. M. Schütz, in Phase Transitions and Critical Phenomena vol. 19, ed. C. Domb and J. L. Lebowitz, London Academic (2001).
39. I. M. Gelfand and S. V. Fomin, *Calculus of Variations* (Dover, New York 2000).